\documentclass[11pt]{elsart}
\usepackage[dvips]{graphicx}

\begin{document}
\begin{frontmatter}

\title{Accelerating consensus of self-driven swarm via adaptive speed}
\author{Jue Zhang$^{a}$ },
\author{Yang Zhao$^{a}$ },
\author{Baomei Tian$^{a}$ },
\author{Liqian Peng$^{a}$ },
\author{Hai-Tao Zhang$^{b,c}$},
\author{Bing-Hong Wang$^{a}$ },
\author{Tao Zhou$^{a,d}$ }
\ead{zhutou@ustc.edu}

\address
{$^{a}$ Department of Modern Physics and Nonlinear Science Center,
University of Science and Technology of China,  Hefei Anhui,
230026, PR China }
\address
{$^{b}$ Department of Engineering, University of Cambridge,
Cambridge CB2 1PZ, U.K.}
\address
{$^{c}$ Department of Control Science and Engineering, Huazhong
University of Science and Technology, Wuhan 430074, PR China}
\address
{$^{d}$ Department of Physics, University of Fribourg, Chemin du
Muse 3, CH-1700 Fribourg, Switzerland}

\begin{abstract}

In resent years, Vicsek model has attracted more and more attention
and been well developed. However, the in-depth analysis on the
convergence time are scarce thus far. In this paper, we study some
certain factors that mainly govern the convergence time of Vicsek
model. By extensively numerical simulations, we find the convergence
time scales in a power law with $r^2\ln N$ in the noise-free case,
where $r$ and $N$ are horizon radius and the number of particles.
Furthermore, to accelerate the convergence, we propose a new model
in which the speed of each particle is variable. The convergence
time can be remarkably shortened compared with the standard Vicsek
model.

\end{abstract}

\date{}
\end{frontmatter}

\section{Introduction}

In nature, collective motion of large numbers of organisms is one of
the most ubiquitous biological phenomena, from motion of groups of
ant \cite{1}, colonies of bacteria \cite{2} and cells \cite{3} in a
small scale, to migration of flocks of birds and schools of fish
\cite{4} in a large
 scale. Those different
forms of collective behaviors can only be understood by considering
the very number of large interactions among group members. Studies
on this issue is significant. On one hand, we can deepen our
understanding of such collective behaviors; on the other hand, we
could probably extract some generic rules from those natural
systems, which can be applied to other relevant realms, such as the
control of unmanned vehicles or robots \cite{5}.

Inspired by biological collective motion, Vicsek\emph{ et al.}
\cite{6} described each individual in the collective motion as an
self-driven particle moving with a constant speed and adjusting the
direction according to the average direction over neighborhood. By
this means, each particle will achieve the same velocity through
finite steps \cite{7}, namely a kinetic phase transition from
disorder to order state \cite{6,8,9}. Besides, many modified models
about self-driven swarm are also proposed. For example, Couzin\emph{
et al.} \cite{10,11} proposed a more biologically realistic model,
and studied the effect of the repulsion, alignment, attraction and
leadership on swarm formation.

In a word, though its simplicity, the Vicsek model is of great
academic significance \cite{12}, which has become a theoretical
templet for the consensus of swarm. However, previous works mostly
focus on the analysis of steady state, while they pay little
attention on the time used to achieve that steady state. As a new
aspect of investigating the collective motion, the convergence time
also has great significance to be explored. On one hand, it can be a
criterion of evaluation of a consensus strategy. On the other hand,
detailed discussions on the relationship between the factors and the
convergence time can provide us some insights about the collective
dynamics. Therefore, in this paper, we will study which factors
influence the convergence time and how can a group of agents get to
be coherent as quickly as possible. First, we will discuss the
relationship between the convergence time and the particle density
as well as the horizon radius of individuals. Then, considering the
fact that the speed of an individual in biological group or
removable robots usually can vary in a certain range, we abandon the
strong assumption in the Vicsek model, namely, the constant speed.
Instead, we propose a swarm model with variable speed, upon which a
new communication protocol is designed, and the convergence time
under this protocol is shorter than that of the standard Vicsek
model.

\section{Convergence time in the Vicsek model}
A group of $N$ particles are considered which are moving in an
$L\times L$ square with periodic boundary conditions. In the Vicsek
model, the particles are moving in an identical constant speed but
different directions, with the initial conditions that the particles
are randomly distributed in the square, and their initial moving
directions are uniformly distributed in the interval $[-\pi,\pi)$.
At each time step, the direction of each particle is determined by
the average direction of the velocity of all the particles
(including itself) within the circle of horizon radius $r$ centered
the given particle. Mathematically speaking, the position of the
$i$th particle is updating according to:
\begin{equation}
\vec{x}_{i}(t+\delta t)=\vec{x}_{i}(t)+\vec{v}_{i}(t)\delta t,
\end{equation}
while the corresponding discrete presentation is:
\begin{equation}
\vec{x}_{i}(t+1)=\vec{x}_{i}(t)+\vec{v}_{i}(t).
\end{equation}
And its direction is updating as:
\begin{equation}
\theta_i(t+1)=\langle\theta_i(t)\rangle_r+\Delta\theta_i,
\end{equation}
where \(\Delta\theta_i\) denotes the noise, and
\(\langle\theta_i(t)\rangle_r\) denotes the average direction of all
the particles within the horizon radius \emph{r}, including the
$i$th particle itself.  \(\Delta\theta_i\) is a random variable
uniformly distributed in the interval $[- \eta /2,\eta /2]$.
Obviously, \(\langle\theta_i(t)\rangle_r\) is given by:
\begin{equation}
tan[\langle\theta_i(t)\rangle_r]=\frac{\langle
v_isin\theta_i(t)\rangle_r}{\langle v_icos\theta_i(t)\rangle_r}.
\end{equation}
\par
Moreover, in order to measure the degree of consensus of all the
particles, an order parameter [6] is introduced as
\begin{equation}
\phi  = \left| {\frac{{\sum\limits_{i = 1}^N {\vec v_i }
}}{{\sum\limits_{i = 1}^N {\left| {\vec v_i } \right|} }}} \right|,
\texttt{  }0\leq \phi \leq 1.
\end{equation}
A larger value of $\phi$ indicates a better consensus, especially
when $\phi=1$, all particles move towards the same direction. In the
circumstances of high density and low noise, all the particles will
definitely approach the consensus state, namely
reach the same direction of velocity after finite time steps \cite{7}. \\[-1cm]
\par
Because of the limitation of horizon radius, each particle can only
communicate with partial particles within the radius and change its
direction according to this local information. Different horizon
radii and particle densities will give rise to diverse convergence
behaviors and times, which will be systematically investigated
later. Without lose of generality, the area is fixed, thus the
particle density can be directly
represented by the number of particles, \emph{N}.\\[-1cm]
\par

\begin{center}
\begin{figure}[!h]
\begin{center}
\includegraphics[scale=1]{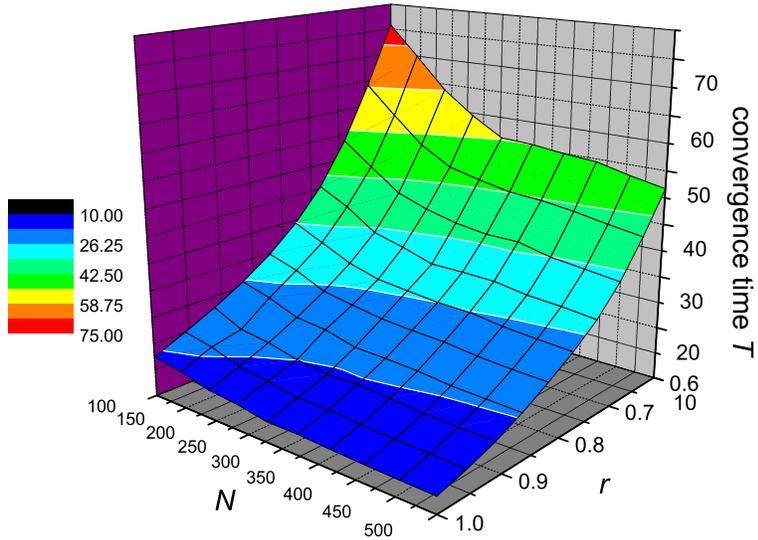}

\end{center}
\caption{(color online) Dependence of convergence time $T$ on
horizon radius $r$ and the number of particles $N$. In the
simulation, all the particles move in a square shaped
 plane of linear size $L=5$, with a constant speed as $v=0.05$ . Control parameters $r$ and $N$,
respectively, vary from 0.5 to 1 and from 100 to 500. The
convergence time is obtained from the average over 500 independent
runs.}
\end{figure}
\end{center}

The simulation results about the convergence time $T$ based on the
noise-free Vicsek model ($\eta$=0) are shown in Fig. 1, where $T$
represents the number of time steps taken before the order parameter
$\phi$ firstly reaches the threshold $\phi_t=0.95$. As long as the
threshold $\phi_t$  is larger enough (close to 1), the variance of
its specific value will not impact much on the qualitative results
shown in this paper. Fig. 1 clearly indicates that the convergence
time decreases with the increase of radius and density. This fact
can be explained in the following two aspects. (a) Given the density
of particles, the larger the horizon radius, the less steps taken to
reach synchronization, because at each time step each particle can
receive more information from others and thus make the adjustment of
velocity more comprehensively, namely more close to the final
convergence velocity, than those with shorter radius. (b) Given
horizon radius, when the number of particles increases, although the
percentage of particles which communicate with a given particle does
not increase as well, higher density is also helpful to reduce the
convergence time, because the particle is making more integral
adjustments at each time step for the increase of the number of
other particles inside its horizon (those particles possess
different directions of velocity, come from different areas a time
step before, bring in information from different areas at latest
time step and then pass the information of current adjusted velocity
to
different areas at the following time step).\\[-1cm]
\par

\begin{center}
\begin{figure}[!h]
\begin{center}
\includegraphics[scale=1]{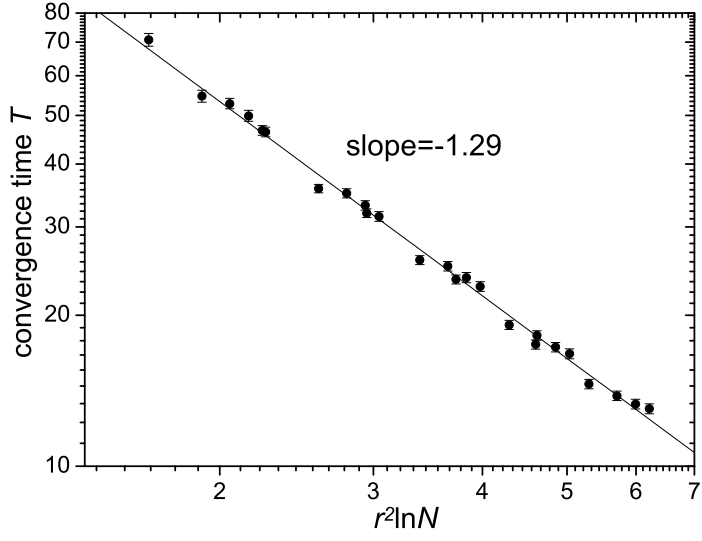}

\end{center}

\caption{ Convergence time $T$ as a function of $r^2\ln N$. The data
points can be well fitted linearly in double logarithmic
coordinates, with slope and error bars marked in the plot. Each data
point is collected from the average over 500 independent runs.}

\end{figure}
\end{center}

\begin{center}
\begin{figure}[!h]
\begin{center}
\includegraphics[scale=1]{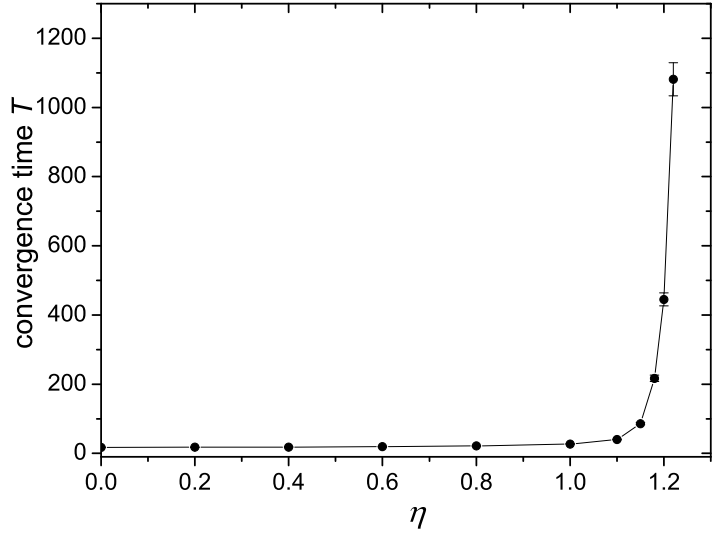}

\end{center}

\caption{ Convergence time $T$ as a function of the noise $\eta$. We
set $r$ =1.0 and $N$=100.}

\end{figure}
\end{center}

Through extensive simulations we eventually come to a numerical
conclusion that the convergence time \emph{T} follows a power law
with $r^2\ln N$. Fig. 2 shows a numerically approximate relation:
\begin{equation}
T\sim{(r^2\ln N)}^{-1.29}.
\end{equation}

In the case with noise, namely $\eta>0$, the convergence time will
increase with the noise strength. In Fig. 3, when $\eta<1.1$, the
convergence time increases slightly, while when $\eta>1.1$, the
convergence time increases dramatically, especially when $\eta>1.2$,
the order parameter can not achieve $\phi_t$=0.95, namely with high
noise this collective motion can not achieve consensus. This result
is quite in according with Fig. 2 in Ref. [6]. Though quantitatively
different, the sharp increasing of $T$ (as the increasing of $\phi$)
can be generally observed for different parameters ($r, N,\phi_t$).

\par
\section{Fast convergence collective motion with variable speed}
Without taking account of generation and annihilation of particles,
the number of particles would be generally fixed. In addition, it
might be difficult to accelerate the consensus process by simply
changing the horizon radius, for which means higher requirement on
both technology and the hardware cost. Considering the speed of
particle in collective motion systems can be variable \cite{13,14},
we propose a new consensus strategy that achieves improvement in
terms of the convergence time compared with the standard Vicsek
model.

With limited horizon radius, a particle could only make the judgment
by itself from the local information it receives. In a completely
chaotic case, although each particle
 updates its
direction to the average one in its local area, this average
direction may be far different from the final synchronization
direction. Therefore, it should be cautious to act with a
comparatively conservative strategy, that is, taking a relatively
lower velocity to prevent from the unnecessary change of its
position. Here the unnecessary change means that, if a particle
changes its position hastily in order to communicate with another
group of particles under such chaotic circumstance, it would be sure
that such impatient behaviors are of no help. On the contrary, if
all the particles choose the conservative strategy of moving with a
comparatively lower velocity, then every particle could have
sufficient time to communicate with its neighbors, leading to a
possibly faster convergence. Only when a certain moving direction is
dominant among its neighbors of a particle, it can surely update its
direction as that one and take a relatively higher velocity since in
that case it is unnecessary for this particle to continue
hesitating. Therefore, a particle's velocity should be somehow
determined by the degree of its local consensus.

In order to present the degree of local synchronization
quantitatively, $\phi_{i}$ is introduced, called the local order
parameter of consensus for the $i$th particle:
\begin{equation}
\phi_{i}= \left| {\frac{{\sum\limits_{j = 1}^{N_i} {\vec v_{ij}  }
}}{{\sum\limits_{j = 1}^{N_i} {\left| {\vec v_{ij}  } \right|} }}}
\right|,\texttt{    }0\leq \phi_{i}\leq 1,
\end{equation}
where $N_i$ is the number of particles within the horizon radius of
the $i$th particle (including itself). The larger value of
$\phi_{i}(t)$, the higher degree of local consensus among the
neighbors of the $i$th particle. Especially when $\phi_{i}(t)=1$,
all $i$'s neighbors move towards the same direction.

Besides, in order to compare with the Vicsek model in terms of the
convergence time, we suppose that the absolute speed can vary from 0
to 0.05. According to the discussion mentioned above, this consensus
strategy with variable speed should satisfy:
\\[-1cm]

a) When all particles in $i$'s neighborhood arrive at an ordered
direction as $\phi_{i}(t)=1$, $v_{i}(t+1)=0.05$;
\\[-1cm]

b) When $\phi_{i}(t)=0$, namely the motions of the particles in
$i$'s neighborhood is completely disorder, $v_{i}(t+1)$ approaches
to zero.

Consequently, we set the speed of the $i$th particle as:
\begin{equation}
v_{i}(t+1)=v_{max}e^{\beta[\phi_{i}(t)-1]},
\end{equation}
where $v_{max}$ is set as 0.05 in this paper. Here $\beta$ is a free
parameter. When $\beta=0$ the protocol degenerates to the standard
Vicsek model, while for $\beta>0$, a particle will move faster in a
better synchronized local circumstance. Indeed, in the present
protocol, speed not only determines the position in the next time
step, but also serves as a carrier, possessing the information about
the local order parameter. The moving direction of the $i$th
particle is also updated following Eq. (4). Note that, when $\phi$
approaches to 1, $\phi_i$ gets close to 1 as well. Therefore from
Eq. (8) we know that the speeds of all particles at this time are
all close to 0.05, which automatically ensure consensus on the
absolute speed.

To sum up, the present protocol with adaptive speed can be described
as
 follows:\\[-1cm]

(1) Determine the initial position and speed of every particle \emph{i};\\[-1cm]

(2) Evaluate the local order parameter of each particle, and
determine its next direction and speed according to Eqs. (4) and (8);\\[-1cm]

(3) Calculate the current order parameter $\phi$ of all particles;\\[-1cm]

(4) Repeat (2) and (3) until $\phi$  approaches to 1.\\
\begin{center}
\begin{figure}[!h]
\begin{center}
\includegraphics[scale=1]{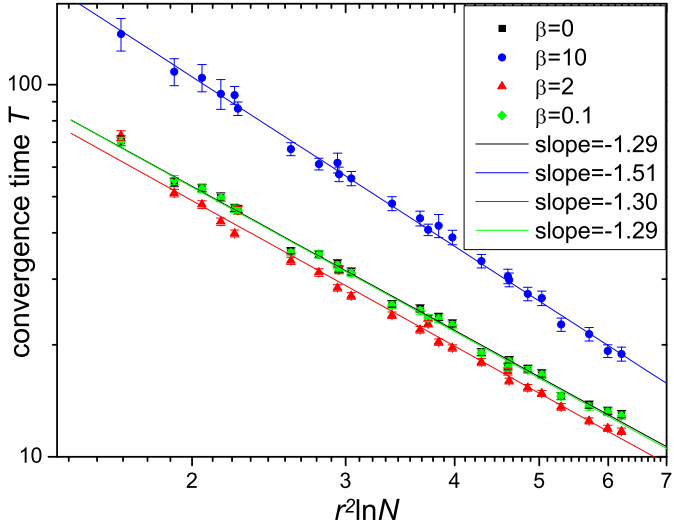}

\end{center}
\caption{(color online) Convergence time $T$ as a function of
$r^2\ln N$ with different $\beta$. The data points can be well
fitted linearly in a double logarithmic coordinates, with slopes and
error bars marked in the plot. Control parameters $r$ and $N$,
respectively, vary from 0.5 to 1 and from 100 to 300. The
convergence time is obtained from the average over 500 independent
runs.}
\end{figure}
\end{center}

In the numerical simulations, we first consider the situation
without noise ($\eta=0$), and still assume that all particles move
in a square-shaped plane of linear size $L=5$. The relation between
the convergence time $T$ and the horizon radius $r$ and the number
of particles $N$ with different $\beta$ is shown in Fig. 4. From
this figure, we can see that in this new protocol the convergence
time $T$ also has a power function with $r^2 \ln N$, the same rule
as illustrated in section 2. Especially, when $\beta$ is not large
enough ($\beta < 5$), the exponents are almost the same, nearly
equal to 1.30. More importantly, in Fig. 4 we can find that,
compared with the standard Vicsek model, the convergence time can be
shortened, which demonstrate the advantage of this modified model in
terms of the convergence time.

\begin{center}
\begin{figure}[!h]
\begin{center}
\includegraphics[scale=1]{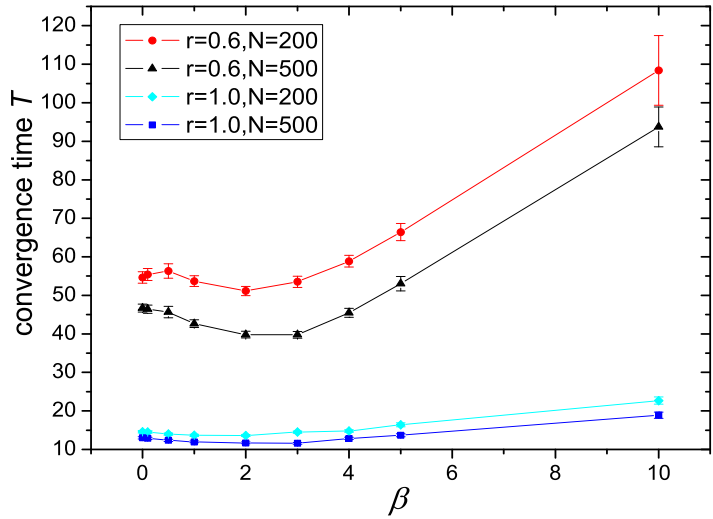}

\end{center}
\caption{(color online) Convergence time \emph{VS}. $\beta$ with
different $r$ and $N$.}
\end{figure}
\end{center}

To further explore this point, we keep the horizon radius $r$ and
the number of particles $N$ fixed and try to find out the optimal
$\beta$ subject to the shortest convergence time. In Fig. 5, we can
see that in all the four curves the optimal $\beta$ is 2 and this
optimal $\beta$ is not susceptive to the horizon $r$ and the number
of particles $N$. Especially, the convergence time can be highly
shortened when the horizon radius $r$ is small.

\begin{center}
\begin{figure}[!h]
\begin{center}
\includegraphics[scale=1]{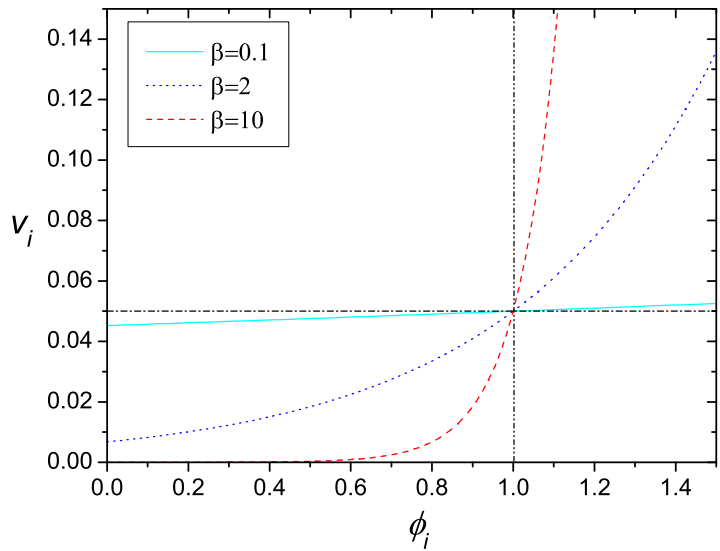}

\end{center}
\caption{(color online) Function
$v_i(\phi_i)=0.05e^{\beta[\phi_i-1]}$ for $\beta$= 0.1, 2 and 10.}
\end{figure}
\end{center}

In Fig. 6, we can see that when $\beta$ is small (e.g. $\beta=0.1$),
no matter what is the current speed distribution, the speeds of most
particles in the next time will approach to the upper limit 0.05,
resulting in almost the same performance as the standard Vicsek
model; besides, when $\beta$ is too large (e.g. $\beta=10$), the
speeds of particles will tend to be 0, thus even when the neighbors
of a given particle is relatively ordered, it still keeps rest and
therefore cannot intercommunicate with others, thus resulting in a
extremely long convergence time. For a proper value of $\beta$ (e.g.
$\beta=2$), when a particle has a large local order parameter, it
will have a great confidence on its current direction, thus leaving
the area with a high speed and using the information (speed
direction) to influence others; Otherwise, it will cast doubt on
whether the current direction is up to the demand of consensus, and
in order to avoid misguiding others with its unconfirmed direction,
it will march on with a relatively slow speed.

\section{Conclusions}
The collective dynamics of intelligent agents is not only the
extensive phenomena in nature, but also an important problem
required in-depth investigation in engineering. Most of the previous
studies concentrated on the depiction and modeling of the swarm
itself. The systematical analysis about the convergence time were
rarely reported. In this paper we have studied the relationship
between the convergence time and the particle density as well as the
horizon radius in the Vicsek model, and found that the convergence
time $T$ has a power function of $r^2\ln N$. Furthermore, we have
designed a motion protocol variable speed, in which the speed is not
only a moving parameter, but also can deliver some information about
its situation of local synchronization. Under such protocol, the
convergence process can be remarkably accelerated compared with the
standard Vicsek model.

\section{Acknowledgements}
We acknowledge Ming Zhao, Luo-Luo Jiang and Da-Jie Zeng for their
assistances for manuscript preparation. This work is partially
supported by the National Natural Science Foundation of China under
Grant Nos. 10472116 and 10635040, as well as the 973 Project
2006CB705500. H.T.Z. acknowledges the National Science Foundation of
China under Grant No. 60704041.

\end{document}